\documentclass{ws-procs9x6-cpt19}

\def\al{\alpha}
\def\be{\beta}
\def\ga{\gamma}
\def\de{\delta}
\def\ep{\epsilon}

\def\ze{\zeta}
\def\et{\eta}
\def\th{\theta}

\def\ka{\kappa}
\def\la{\lambda}

\def\rh{\rho}

\def\si{\sigma}

\def\Ga{\Gamma}

\def\cG{{\cal G}}

\def\cL{{\cal L}}

\def\fr#1#2{{{#1} \over {#2}}}

\def\frac#1#2{{\textstyle{{#1}\over {#2}}}}

\def\prt{\partial}

\def\pt#1{\phantom{#1}}

\newcommand{\beq}{\begin{equation}}
\newcommand{\eeq}{\end{equation}}
\newcommand{\bea}{\begin{eqnarray}}
\newcommand{\eea}{\end{eqnarray}}
\newcommand{\bit}{\begin{itemize}}
\newcommand{\eit}{\end{itemize}}
\newcommand{\rf}[1]{(\ref{#1})}

\def\mn{{\mu\nu}}

\def\sb{\overline{s}}

\def\st{\tilde{s}}

\begin{document}

\newcommand{\refeq}[1]{(\ref{#1})}
\def\etal {{\it et al.}}
%any other macros go here 

\title{Recent Developments in Spacetime-Symmetry tests in Gravity}

\author{Q.G.\ Bailey}

\address{Embry-Riddle Aeronautical University,\\
Prescott, AZ 86301, USA}

\begin{abstract}
We summarize theoretical and experimental work on tests
of CPT and local Lorentz symmetry in gravity.
Recent developments include extending the effective field theory framework 
into the nonlinear regime of gravity.

\end{abstract}

\bodymatter

\section{Introduction}

Motivated by potentially detectable but minuscule signatures 
from Planck-scale or other new physics, 
there has been a substantial increase 
in tests of spacetime symmetry in gravity in recent years.\cite{string,datatables}
Some novel hypothetical effects that break local Lorentz symmetry and CPT symmetry in gravitational experiments as well as 
solar system and astrophysical observations have been studied in recent works.\cite{review}  
Much of this work uses the effective field theory framework, 
the Standard-Model Extension (SME), 
that includes gravitational couplings.\cite{sme1,sme2}  
In other cases, 
the parameters in specific hypothetical models 
of Lorentz violation in gravity have been tested.\cite{vecCS}

\section{Framework}

The general framework of the SME in the pure-gravity sector 
can be realized as the Einstein-Hilbert action plus a series
of terms formed from indexed coefficients, 
explicit or dynamical, 
contracted with increasing powers of curvature and torsion.
Each term in this series
maintains observer invariance of physics, 
while breaking ``particle" invariance, 
with respect to local Lorentz symmetry and diffeomorphism symmetry.\cite{sme2}

One interesting and practical subset of the SME 
is a general description of CPT and Lorentz violation 
that is provided by an expansion valid for linearized gravity
($g_\mn = \et_\mn + h_\mn$).  
For instance, 
in this approximation the Lagrange density for General Relativity (GR)
plus the mass dimension 4 and 5 operators 
controlling local Lorentz and CPT violation are given by\cite{bk06,bkx15,km16}
\beq
\cL=-\frac {1}{4\ka} (h^\mn G_\mn - \sb^{\mu\ka} h^{\nu\la} \cG_{\mu\nu\ka\la}
+\frac {1}{4} h_\mn (q^{(5)})^{\mu\rh\al\nu\be\si\ga} \prt_\be R_{\rh\al\si\ga}+...), 
\label{L}
\eeq
where $\ka=8\pi G_N$, 
and the double dual curvature $\cG$ 
and the Riemann curvature $R_{\rh\al\si\ga}$ are linearized in $h_\mn$.
This lagrange density maintains linearized diffeomorphism invariance, 
though generalizations exist\cite{km18}, 
and $\sb_\mn$ and $(q^{(5)})^{\mu\rh\al\nu\be\si\ga}$ are the coefficients controlling 
the degree of symmetry breaking (they are zero in GR).

\section{Experiment and Observation}

The mass dimension 4 Lagrange density, 
the minimal gravity SME, 
has now been studied in a plethora of tests. 
The best controlled and simultaneous parameter-fitting limits 
come from lunar laser ranging\cite{llr}, 
and other laboratory experiments such as gravimetry.\cite{gravi}
These place limits on the $\sb_\mn$ coefficients at the level
of approximately $10^{-7}-10^{-8}$ on the 3 $\sb_{TJ}$ and $10^{-10}-10^{-11}$ on 5 of the $\sb_{JK}$ coefficients.
Stronger limits can be countenanced from distant cosmic rays\cite{kt15},
and one combination of coefficients is bounded at $10^{-15}$
by the multimessenger neutron star inspiral event in 2017.\cite{GRBGW} 
Other searches for these coefficients include ones with pulsars.\cite{shao}

For the mass dimension 5 coefficients in \rf{L} that 
break CPT symmetry, 
the post-Newtonian phenomenology includes a velocity-dependent inverse cubic force. 
This leads to an extra term in the relative acceleration of two bodies given by\cite{bh17}
\bea
\de a^j &=&\fr {G_N M v^k}{r^3} 
\big( 15 n^l n^m n^n n_{[ j } K_{k] lmn} 
\nonumber\\
&&
+  9 n^l n^m K_{[jk] lm}-9 n_{[j} K_{k] ll m} n^m -3 K_{[jk]ll} \big),
\label{de_acc}
\eea
where $K_{jklm}$ are linear combinations of the coefficients $q$ in the lagrange density \rf{L}, 
$\vec r$ is the separation between the bodies and ${\hat n}={\vec r}/r$.

Measurements of the mass dimension $5$ coefficients in \rf{de_acc} 
are currently scarce.  
There is one constraint on a combination of dimension 5 and 6 coefficients from Ref.\ \refcite{km16} in searches for dispersion of gravitational waves from distant sources
and analysis with multiple gravitational wave events 
is underway.\cite{gwfit}
Disentangled constraints on the $K_{jklm}$ coefficients from analysis of pulsar observations exists at the level of $10^{6}$ meters.\cite{shao2}
This leaves room for potentially large, 
``countershaded" symmetry breaking to exist in nature.\cite{kt09}
Higher-order terms in the series, at mass dimension $6$ and beyond, 
have been constrained in short-range gravity tests.\cite{sr}

\section{Extension to the nonlinear regime}

While the general form for linearized gravity has been explored, 
only several works have explored the general SME framework beyond linearized gravity.\cite{nl}  
One approach is to extend the general lagrange density for linearized gravity 
(which is quadratic order in the metric fluctuations)
to include terms of cubic and higher order terms.  
If we adopt the point of view of spontaneous symmetry breaking (SSB), 
one must consider the dynamics of the coefficients for Lorentz violation.  
Considering the case of a symmetric two tensor $s_\mn$ being the 
Lorentz-breaking field, 
it is expanded in the SSB scenario as $s_\mn = \sb_\mn + \st_\mn$, 
where $\sb_\mn$ are the vacuum expectation values and $\st_\mn$ are the fluctuations.
The Lagrange density is a series $\cL = \cL^{(2)} + \cL^{(3)}+...$
where $(2)$ and $(3)$ indicate the order in fluctuations $h_\mn$ or $\st_\mn$.
A general conservation law\cite{bluhm15}, 
contained in equation (9) 
of Ref.\ \refcite{bluhm16}, 
can be used to constrain the terms in the series.
In the example of $s_\mn$ it takes the form
\beq
\prt_\be \left( \fr {\de \cL}{\de h_{\ga\be}} \right) 
+ \Ga^\ga_{\pt{\be}\al\be} \left(\fr {\de \cL}{\de h_{\al\be}} \right) 
+ g^{\de\ga} s_{\de\al} \prt_\be \left( \fr {\de \cL}{\de \st_{\al\be}} \right) 
+ g^{\de\ga} {\tilde \Ga}_{\de\al\be} \fr {\de \cL}{\de h_{\al\be}}=0,
\label{constr}
\eeq
where
${\tilde \Ga}_{\de\al\be}=
(\prt_\al \st_{\be\de}+\prt_\be \st_{\al\de}-\prt_\de \st_{\al\be})/2$.
This equation holds ``off-shell", 
assuming the action obtained from $\cL$ is diffeomorphism invariant.

In the case of the minimal SME with just $s_\mn$, 
the Lagrange density is constructed from all possible contractions of generic terms of the quadratic form $\sb_{\al\be} h_{\ga\de} \prt_\ep \prt_\ze h_{\et\th}, 
\st_{\al\be} \prt_{\ga} \prt_\de \st_{\ep\ze}, 
\st_{\al\be} \prt_\ga \prt_\de h_{\ep\ze}, ...$, 
the cubic form 
$\sb_{\al\be} h_{\ga\de} h_{\ep\ze} \prt_\et \prt_\th h_{\ka\la}, 
\sb_{\al\be} h_{\ga\de} \prt_\ep h_{\ze \et} \prt_\th h_{\ka\la}, 
h_{\al\be} \st_{\ga\de} \prt_\ep \prt_\ze \st_{\th\ka},...$, 
and potential terms.
The sum of all such terms, each with an arbitrary parameter, 
is inserted into \rf{constr} and the resulting linear equations
for the parameters are solved.
What remains, 
up to total derivative terms in the action, 
are a set of independently diffeomorphism invariant terms.
As an example of such a term produced by this expansion, 
we find to cubic order
\bea
\cL &\supset& \sb_{\al\be} \st^{\al\be} R^{(1)} +
\frac 12 \st_{\al\be} \st^{\al\be} R^{(1)} 
-2 h^{\al\be} \sb_\al^{\pt{\al}\ga} \st_{\be\ga} R^{(1)} 
\nonumber\\
&&
+\sb_{\al\be} \st^{\al\be} 
(\Ga_{\ga\de\ep} \Ga^{\ga\de\ep}  
- \Ga^{\ga\de}_{\pt{\ga\de}\de} \Ga_{\ga\ep}^{\pt{\ga\ep}\ep} 
+ \frac 12 h^\ga_{\pt{\ga}\ga} R^{(1)} -2 h^{\ga\de} R^{(1)}_{\ga\de}),
\label{sample}
\eea
where the $(1)$ superscript indicates linear order in $h_\mn$ and
the connection coefficients are at linear order.
Note that this construction generally includes dynamical terms for the 
fluctuations and so does not assume ``decoupling".\cite{ms}

The construction including all such terms allows exploration of the regime in gravity where nonlinearities need to be considered.\cite{b19,b16}
This includes higher order post-Newtonian gravity in weak-field systems and developing a multipole expansion for gravitational waves affected by Lorentz violation.

\section*{Acknowledgments}
This work was supported by the National Science Foundation 
under grant no.\ 1806871 and Embry-Riddle Aeronautical University's 
FIRST grant program.

\end{document}